\documentclass[apl,superscriptaddress,reprint,twocolumn,showpacs]{revtex4-1}

\usepackage{amsfonts}

\usepackage{graphicx}
\usepackage{epstopdf}
\usepackage{amsmath}
\usepackage{amssymb}
\usepackage{verbatim}
\usepackage[ulem=normalem]{changes}
\usepackage{booktabs}
\usepackage{natbib}

\usepackage{xcolor}
\usepackage{gensymb}
\usepackage{xcolor}

\newcommand\crule[3][black]{\textcolor{#1}{\rule{#2}{#3}}}
\usepackage{tikz}

\begin{document}
\title{Cooling and Manipulation of a levitated nanoparticle with an optical fiber trap}

\author{Pau Mestres}
\altaffiliation{These authors equally contributed to this work}
\author{Johann Berthelot}
\altaffiliation{These authors equally contributed to this work}
\affiliation{ICFO-Institut de Ciencies Fotoniques, The Barcelona Institute of Science and Technology, 08860 Castelldefels (Barcelona), Spain}
\author{Marko Spasenovi\'{c}}
\altaffiliation{Current address: Center for Solid State Physics and New Materials, Institute of Physics Belgrade, University of Belgrade, Pregrevica 118, 11080 Belgrade, Serbia.}
\affiliation{ICFO-Institut de Ciencies Fotoniques, The Barcelona Institute of Science and Technology, 08860 Castelldefels (Barcelona), Spain}
\author{Jan Gieseler}
\altaffiliation{Current address: Physics Department, Harvard University, Cambridge, MA 02318, USA}
\affiliation{Photonics Laboratory, ETH Zurich , 8093 Zurich, Switzerland}
\author{ Lukas Novotny}
\affiliation{Photonics Laboratory, ETH Zurich , 8093 Zurich, Switzerland}
\author{ Romain Quidant}
\affiliation{ICFO-Institut de Ciencies Fotoniques, The Barcelona Institute of Science and Technology, 08860 Castelldefels (Barcelona), Spain}
\affiliation{ICREA $-$ Instuci\'o Catalana de Recerca i Estudis Avan\c cats, 08010 Barcelona, Spain.}

\begin{abstract}
Accurate delivery of small targets in high vacuum is a pivotal task in many branches of science and technology. Beyond the different strategies developed for atoms, proteins, macroscopic clusters and pellets, the manipulation of neutral particles over macroscopic distances still poses a formidable challenge. Here we report an approach based on a mobile optical trap operated under feedback control that enables cooling and long range 3D manipulation of a silica nanoparticle in high vacuum. We apply this technique to load a single nanoparticle into a high-finesse optical cavity through a load-lock vacuum system. We foresee our scheme to benefit the field of optomechanics with levitating nano-objects as well as ultrasensitive detection and monitoring.
\end{abstract}

\maketitle

Controlled manipulation of matter over long distances in high vacuum has enabled many groundbreaking experiments, including protein characterization via x-ray diffraction \cite{Chapman:2011ei}, macroscopic quantum interference \cite{Hornberger2012Colloquium,RomeroIsart2011Superpositions}, laser induced fusion \cite{Nakai2004fusion}, and quantum information processing \cite{Kuhr:2001SingleAtom}. The sample size ranges from sub-nanometer for single atoms \cite{Kuhr:2001SingleAtom} up to a few millimeters for laser fusion pellets. To avoid unwanted interactions, microscopic samples can be confined in levitation using Paul traps \cite{paul1990electromagnetic} or optical fields \cite{ritsch2013cold}. These potentials usually extend over small regions on the order of few microns. Therefore,  a delivery method is needed in order to position the sample of interest into the region of stable potential.

 Atoms and small molecules on a substrate can be delivered into the gas phase through heating, since the additional thermal energy is sufficient to overcome the potential barrier caused by the attractive Van der Waals interaction between the sample and the substrate.
In contrast, larger objects remain stuck because the Van der Waals interaction increases with size. For micron sized objects, the interaction can be overcome mechanically by shaking the substrate with a piezoelectric transducer \cite{ashkin:1971highvac,li2011millikelvin}, allowing the particle to build up enough kinetic energy to escape the interaction potential. However, as this approach relies on the object's mass, it cannot be applied to particles smaller than $\approx 1\mu m$.

 The advent of electrospray ionization partially closed this gap by delivering macromolecules and nanoparticles in solution into the gas phase \cite{Fenn1989Electrospray}. Electrospray ionization generates fast highly charged particles, which are then guided with electrical forces into the  vacuum chamber \cite{paul1990electromagnetic,kuhlicke2015demand}. However, charged particles are not always desirable and  neutral particles cannot be manipulated by electrical forces. 

An approach in the context of recent optomechanical experiments with an optically trapped nanoparticle, uses a nebulizer to deliver neutral particles from solution into the gas phase \cite{ashkin:1971highvac,gieseler2012subkelvin, summers:2008aerosol}. Neutral particles are trapped under ambient conditions and the pressure is reduced while the particle is trapped. Despite its simplicity, this approach is not ideal since it compromises the ultimate vacuum and excess particles degrade the performance of ultra-high reflectivity mirrors, which are required for cavity optomechanics \cite{RomeroIsart2010livingorganisms, chang2010cavity,kiesel2013cavity}. \\

To maintain ultra clean conditions it becomes necessary to physically separate the particle injection region from the experiment and introduce a transport mechanism between them.  Conventional optical manipulation techniques based on tracking interference patterns or non-diffracting beams \cite{cizmar:2005conveyor} are limited to travel ranges below $1\ \rm mm$, which is insufficient to cover the distance between adjacent vacuum chambers. Hollow core optical fibers \cite{schmidt:2012optothermal, schmidt:2012metrology} can overcome this limitation. Yet, their use for fine delivery in high vacuum is still to be demonstrated \cite{grass2013fiber}. Additionally, if no other feedback stabilization mechanisms are used, optically trapped particles in vacuum conditions escape the trap below $\sim$1mBar due to a poorly understood mechanism that has been reported by several groups\cite{millen2014nanoscale,kiesel2013cavity,price2015vacuo}, thus limiting the pressure range in which experiments can be performed. 

In this letter we address these difficulties by using a load-lock scheme with a mobile trap. Particles are loaded under ambient pressure in a first vacuum chamber and then transferred under vacuum into a high finesse cavity inside a second vacuum chamber. Additionally, our trap integrates a detection scheme that follows the particle's motion over arbitrarily long manipulation distances, providing the information required to implement a particle stabilization mechanism. This adds to our high vacuum trapping capabilities the ability to manipulate levitated nano-objects over large distances.

\begin{figure}
 \includegraphics[width=8.5cm]{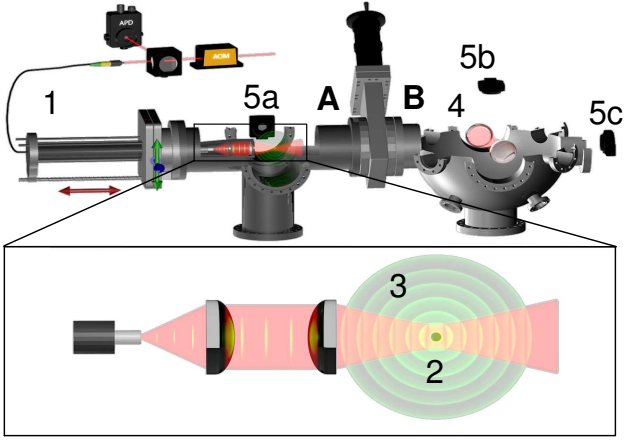}
 \caption{Schematic of the experimental setup: A single mode optical fiber (1) is mounted on a 3D translational stage that goes all the way through 2 vacuum chambers (A and B). In the first chamber (A) the propagating light (red beam) is collimated and focused into a diffraction limited spot (2) using two aspheric lenses. A small dielectric nanoparticle is trapped at the focus (2).  The light scattered from the particle  (green waves in 3) is collected back into the fiber with the same aspheric lenses and interferes with a back-reflection from the vacuum side fiber's facet. The interferometric signal is detected with an APD outside the vacuum chamber.  The second vacuum chamber (B) contains a high Finesse Cavity (4) and is always kept in vacuum. CCD cameras (5a, 5b and 5c) are placed at different viewports for accurate positioning of the trap in the different vacuum chambers.}
 \label{fig: figure1}
\end{figure}

Figure~\ref{fig: figure1} shows the experimental setup. The mobile optical trap (MobOT) consists of an optical fiber, a collimator and a high numerical aperture aspheric lens (NA=0.8), which are mounted together on a metal rod inside our vacuum system. The metal rod is attached to a translation stage equipped with three stepper motors. The stepper motors allow us to move the MobOT in all three spatial directions with 2$\mu m$/step. The step size is given by the ratio of the gears used to actuate the translation stage and can be reduced  on detriment of speed.
Infrared light ($\lambda = 1064 \rm nm$) from a free space laser is coupled into the fiber and sent through the fiber into the vacuum chamber, where it is out-coupled with the collimator and focused by the aspheric lens to form a stable optical trap $1.5$mm away from the surface of the lens.
As in previous experiments \cite{gieseler2012subkelvin,gieseler2014nonlinear} we use a nebulizer to load the optical trap with a single $\rm SiO_2$ nanoparticle with radius $r\sim 75\rm nm$ .\\



The particle scatters light from the trapping beam.  The backscattered light  is  collected with the same aspheric lens used for focusing, coupled back into the fiber via the collimator and sent to an avalanche photodiode (APD). At the APD, the backscattered light from the nanoparticle interferes with laser light that is reflected at the end facet of the vacuum side fiber, acting as a reference arm in a homodyne detection scheme. The intensity at the APD is, thus, given by
\begin{equation}\label{eq:detector_signal}
E_{det}^2=E_r^2+E_p^2+2E_{p}E_r \cos (\phi_r+\phi_p(z))
\end{equation} 
where  $E_p$ ($\phi_p$) and $E_r$ ($\phi_{r}$) are the electric field amplitude (phase) of the backscattered light from the particle and reference, respectively.\\

The radiation pattern of a subwavelength dielectric particle is well approximated by a dipole and, to lowest order, the position of the particle along the optical axis is imprinted only onto the phase of the backscattered light $\phi_p(z)\approx \phi_0 + \phi_z$. Here, $\phi_0$ is an arbitrary relative phase between a stationary particle in the trap center and the reference beam, and $\phi_z = 2 k_0z$ is the phase change of the particle due to its motion along the optical axis, where $k_0 = 2\pi/\lambda$ is the wavevector. Note that a small displacement of $z(t)$ along the optical axis, leads to 2$z(t)$ optical path difference between the backscattered and the reference light.  
For a periodic motion $z(t) = q_z\cos(\Omega_z t)$ the last term of equation~\eqref{eq:detector_signal} reads 
\begin{equation}\label{eq:interference}
2E_p E_r\left( J_0(2k_o q_z)+{\rm Re}\left\{2 \sum^{\infty}_{n=1}e^{i \phi_0}i^n J_n(2 k_0 q_z) \cos(n \Omega_z t)\right\}\right)
\end{equation} 
where we used the Jacobi-Anger expansion \cite{abramowitz1964handbook}.

Thus, the  right term of equation~\eqref{eq:interference} is a sum of harmonics of the particle oscillation frequency where the relative strength of each harmonic is given by a Bessel function  $J_n (2 k_0 q_z)$.”\\

\begin{figure}[h!]
 \includegraphics[width=8.5 cm]{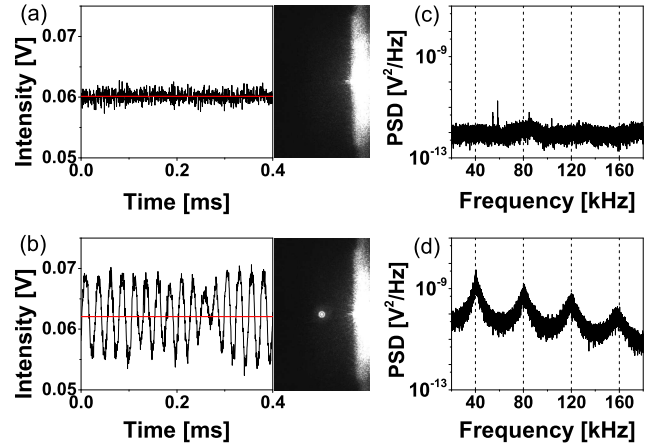}
 \caption{Trapping of a single 75nm radius silica nanoparticle in the MobOT:  Time traces of the interference intensity measured without  (a) and  with (b) a particle in the trap. The red lines in (a) and (b) correspond to the mean level of the signal. From longer time traces we compute the power spectral density for a trap without (c) and with (d) particle. When the particle is present, the PSD shows peaks at the successive harmonics of $\Omega_z$ (grid dashed lines in (d)). The experiment is performed at $2\,\rm mBar$. Insets show a side view of the optical fiber trap in the loading chamber (camera 5a Fig. \ref{fig: figure1}).}
 \label{fig: figure2}
\end{figure}

Figure \ref{fig: figure2} shows a characteristic time trace and power spectral density (PSD) of the detector signal measured with a nanoparticle trapped at $2\rm mBar$. We clearly resolve the oscillatory underdamped motion \cite{Li:2012instantaneous} and observe up to four harmonics of the fundamental frequency at $\Omega_z/2\pi = 40\rm kHz$. From the ratio of the $1^{st}$ and $3^{rd}$ harmonic $J_1(2k_0q_z)^2/J_3(2k_0q_z)^2$ we retrieve $q_z\approx 183\rm nm$ in excelent agreement with what is expected (191nm) from the equipartition theorem $2 T_0 k_B = m \Omega_z^2 q_z^2$, where $T_0 = 300\rm K$ and $k_B$ is Boltzmann's constant, $m = \frac{4}{3}\pi r^3\rho$, $r=73\pm 4\rm nm$ and $\rho = 2200 \rm kg/m^3$.
In addition to the particle oscillations, the APD signal in presence of the particle (Fig.~\ref{fig: figure2}b)differs from the APD signal without a particle (Fig.~\ref{fig: figure2}a)) by an offset. This offset originates from the terms $E_p^2$ and $2E_p E_r J_0 (2k_o q_z)$ in equation~\eqref{eq:detector_signal} and equation~\eqref{eq:interference}, respectively. Remarkably, our detection scheme enables us to follow the particle motion over arbitrary distances. This is in stark contrast to common stationary detection methods, where the detection range is of the order of the particle diameter for quadrant photo detectors (QPDs) based methods \cite{martinez2012back} or limited by the field of view for camera based detection \cite{millen2015cavity,millen2014nanoscale}.




The detection signal is used to feedback stabilize the particle in the MobOT at pressures below 1mBar. In contrast to  previous schemes \cite{li2011millikelvin, ashkin:1977Feedback,gieseler2012subkelvin}, our feedback does not depend on a precise phase relationship between the particle motion and the parametric feedback force. Instead, it measures abrupt changes in the detector signal and penalizes large amplitude oscillations by increasing the trap stiffness akin to a plasmonic self-induced back action trap \cite{juan:2009back-action}\\

\begin{figure}[t!]
 \includegraphics[width=8.5 cm]{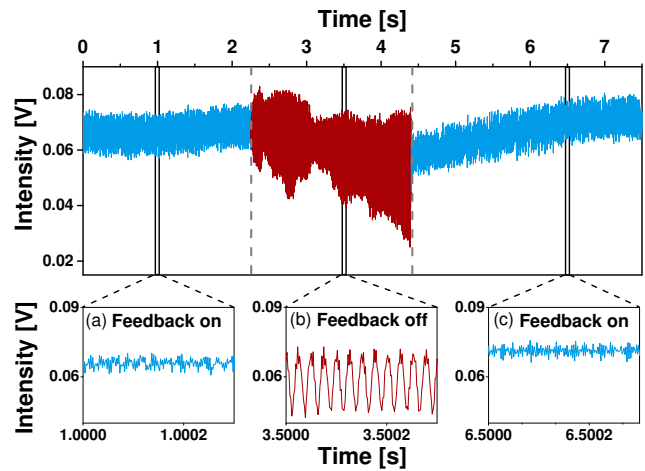}
 \caption{Effect of the feedback on the particle motion (P=$2\times 10^{-5} \rm mBar $): (a) The particle motion amplitude is damped by the feedback. (b) It increases when the feedback is deactivated and (c) returns to a low amplitude oscillation when the feedback is switched on again.}
 \label{fig: figure3}
\end{figure}

The oscillation amplitude of the trapped particle changes randomly due to stochastic collisions between the particle and residual air molecules. The timescale  $\sim 1/\Gamma_0$ over which the amplitude changes is determined by the damping coefficient  $\Gamma_0$ \cite{gieseler2014dynamic}. The damping coefficient depends linearly on the gas pressure $P_{\rm gas}$  \cite{gieseler2012subkelvin} and is much smaller than the oscillation frequency under vacuum conditions (underdamped regime) \cite{Beresnev1990Motion}. The relatively slow change in amplitude allows us to implement the feedback in a FPGA. We sample the detector signal at $520\, \rm kSamples/s$. For each sample $V_i$ we measure its deviation from the current mean $\langle V_z\rangle = M^{-1}\sum_{i-M}^{i-1} V_i$ and compare it to the standard deviation $\sigma_V =\sqrt{M^{-1}\sum_{i-M}^{i-1} (\langle V_z\rangle-V_i)^2}$ over the last $M = 13$ samples, which corresponds to one oscillation period.
To cool the particle, we modulate the laser intensity with an acousto-optic modulator (AOM) according to
\begin{equation}
  P_{\rm Laser} = P_{\rm 0}  \times
  	\begin{cases} 
  		(1+\epsilon)  & \mbox{if } |\langle V_z\rangle - V_z| > \sigma_V \\ 
		1 	& \mbox{if } |\langle V_z\rangle - V_z| \leq \sigma_V
	\end{cases} ,
\end{equation}
where $P_0$ is the laser intensity without feedback and $\epsilon = 7.5\%$ is the laser modulation depth.
Since the trap stiffness $k_{\rm trap}\propto P_{\rm Laser} $, this scheme penalizes large oscillation amplitudes by stiffening the trap whenever the amplitude becomes too large, leading to an overall reduction in the particle oscillation amplitude as shown in Fig.~\ref{fig: figure3}.\\

\begin{figure}[b!]
 \includegraphics[width=8.5cm]{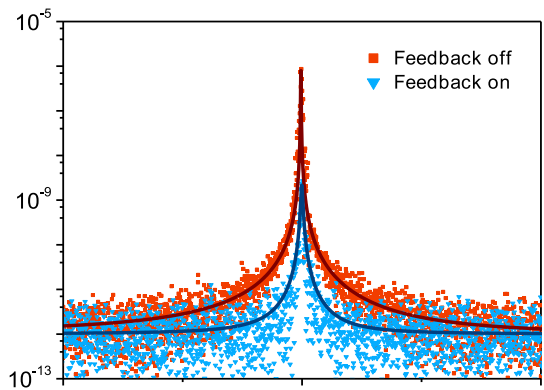}
 \caption{Cooling of the particle motion: Closeup of the PSD in the region of the first harmonic at 40kHz of a particle at $2\times 10^{-5}\,\rm mBar$ without ( \crule[red!100!white!100]{0.25cm}{0.25cm} ) and with feedback ( \textcolor{cyan}{$\blacktriangledown$} ). Darker lines are Lorentzian fits. }
 \label{fig: figure4}
\end{figure}

According to the equipartition theorem, a static increase of the laser power $P_{\rm laser}\propto k_{trap}$ decreases the oscillation amplitude $\langle z^2\rangle \propto 1/P_{\rm laser}$, while increasing the oscillation frequency $\Omega_z\propto \sqrt{P_{\rm laser}}$.
Fig.~\ref{fig: figure4} shows the PSD of the detector signal with and without feedback. The oscillation frequency shifts only slightly to higher frequencies under feedback. This demonstrates that our feedback increases the average laser power only very little. The feedback damping also manifests itself as an $\approx$ 8 fold increase in the resonance linewidth. According to Gieseler et al.\cite{gieseler2012subkelvin}, we can estimate the center of mass temperature ($T_{CM}$) from the additional optical damping as:
\begin{equation}
T_{CM}=\dfrac{T\Gamma_o}{\Gamma_o +\delta \Gamma}
\end{equation} 
Where $T$ is the temperature of the phonon thermal bath, $\Gamma_o$  its dissipation and $\delta \Gamma$ the additional dissipation rate introduced by the optical feedback. This corresponds to $T_{CM}\approx$ 40K. It is worth noticing that due to the nonlinearity of the Bessel functions in equation (\ref{eq:interference}) and especially the nonlinear broadening at low pressures  without feedback \cite{gieseler2013thermal}, we might overestimate $\Gamma_o$ with respect to $\delta \Gamma$, leading to an underestimate of the cooling efficiency. This trend is corroborated by comparing the variance of the signal traces, which gives $T_{CM}\approx30K$.\\

\begin{figure}[h!]
 \includegraphics[width=8.5cm]{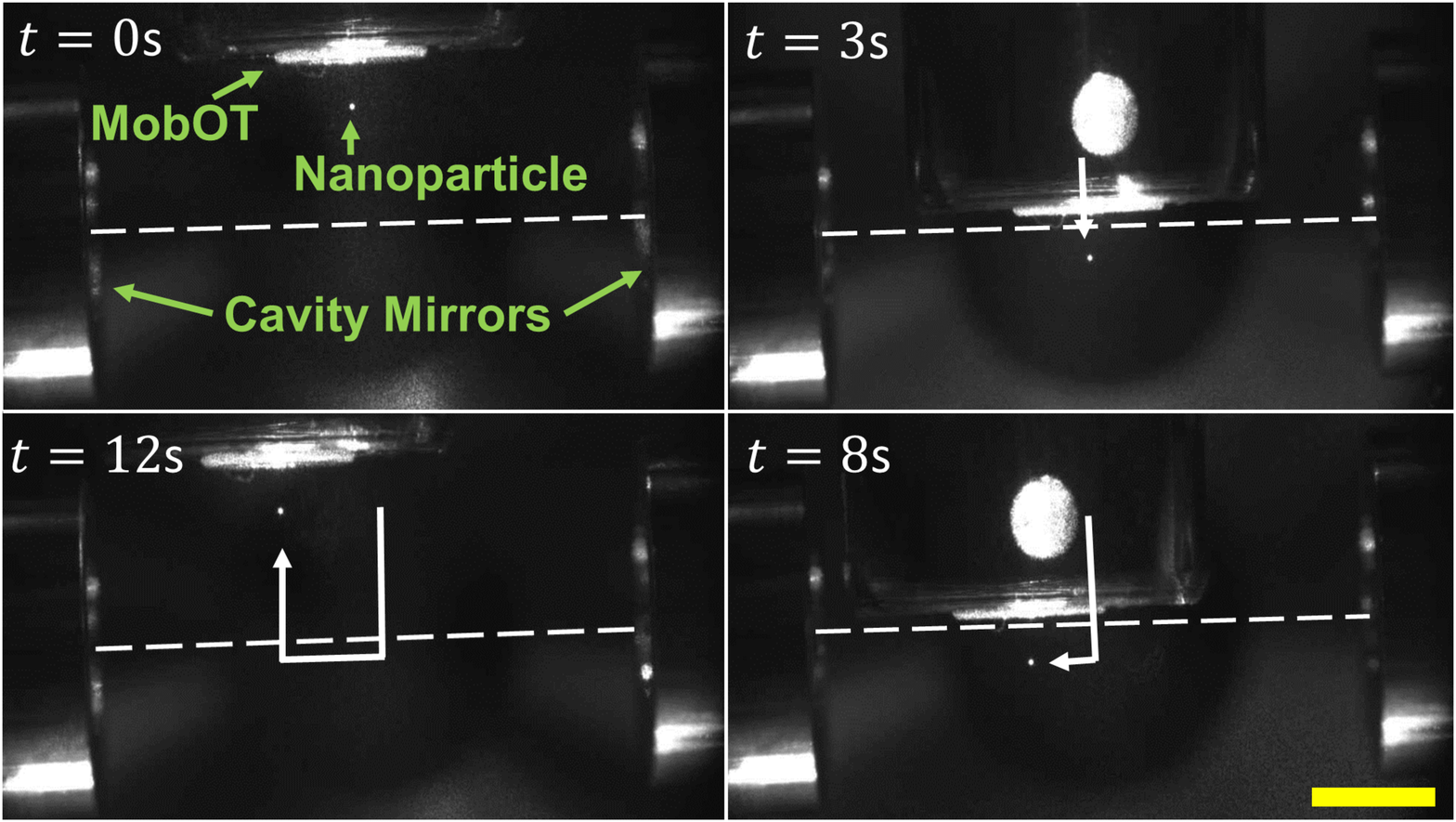}
 \caption{3D fine displacements of a levitated nanoparticle through a high finesse optical cavity: Frames of a record showing the position of the particle and the MobOT while manipulating the trap around the optical axis of the cavity (white dashed line). The solid white line follows the trajectory of the particle and the times are shown in second at the top-left corner of each frame. The yellow scale bar corresponds to a length of 5 mm. The record is made using camera 5b (Fig.~\ref{fig: figure1}), (Multimedia view).}
 \label{fig: figure5}
\end{figure}

We use the MobOT to stabilize a nanoparticle and to translate it between two vacuum chambers, which are separated by a valve and $\approx 66\rm\, cm$ apart (A and B Fig.\ref{fig: figure1}).
First, we trap a particle in chamber A under ambient pressure, while we keep chamber B at $10^{-5}\rm mBar$. We then reduce the pressure in chamber A to $1\rm mBar$ and activate the feedback. 
When chamber A has reached $10^{-4}\rm mBar$  we slowly open the valve. This equilibrates the pressure between the two chambers at $6 \times 10^{-5}\rm mBar$.
Once the equilibrium pressure has been reached, we push the metal rod into chamber B at an average speed of $2\,\rm mm/s$, which takes about 6 min \cite{video1}.
At the center of the second vacuum chamber, which hosts a high finesse optical cavity  (F$\approx 120.000$ estimated by ringdown measurements), we observe the MobOT through the top window using an external camera  (camera 5b in Fig.~\ref{fig: figure1}).
Fig.~(\ref{fig: figure5})(Multimedia view) shows a sequence of camera images as we translate the trapped particle in the $XZ$ plane, demonstrating our excellent control of the particle position throughout the cavity.

\begin{figure}[h!]
 \includegraphics[width=8.5cm]{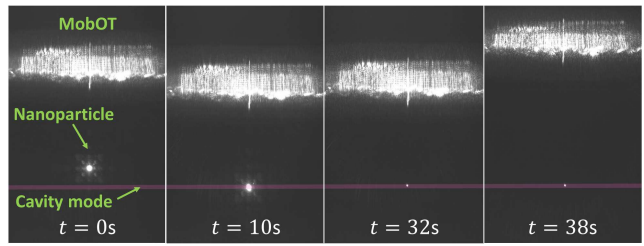}
 \caption{Particle transfer from the MobOT to the cavity field: Frames of a record showing the transfer of a 75nm radius particle from the MobOT to the standing wave of an optical cavity. The diameter of the cavity mode is approximately 130$\mu$m, and has been colored for clarity. The record is made using a magnified view from camera 5b (Fig.~\ref{fig: figure1}),(Multimedia view).}
 \label{fig: figure6}
\end{figure}

Finally, we use the MobOT to transfer the particle into the cavity field as shown in Fig~.\ref{fig: figure6}(Multimedia view). From time $t$=0s to $t=32s$ we position the particle to the cavity mode (65$\mu$m waist) by fine tuning the position of the trap into the three axis. Then, at $t=32s$ we increase the intracavity power to 70W, allowing the particle to jump from one trap to the other. At this point we retract the MobOT to check that the particle is kept traped by the cavity field. Few seconds after the transfer, the particle escapes the cavity field as reported by other groups \cite{kiesel2013cavity}.\\

In conclusion, we have demonstrated position detection of a levitated particle collecting the backscattered light through a fibre. In combination with a mobile optical trap under feedback and a load-lock scheme, it enabled us to translate a levitated particle over long distances and to accurately position it in three dimensions allowing transfer between different trapping potentials even under high vacuum conditions. Our method could be used to deliver nano-objects of different sizes as long as they experience trap depths larger than 10$k_bT$, \cite{ashkin1986observation}. Using the dipole approximation, we estimated a trap depth for the MobOT  of $\approx 30k_b T$. Hence, using the same power we should be able to deliver particles with sizes down to 50nm radius (limited by the particle polarizability). An unpper size bound appears when the particle is larger than the MobOT spot size, therfore we should not be able to trap and deliver particles larger than $\approx 1\mu$m diameter.

 We envision that our approach will enable many exciting experiments that require to deliver a nanometer sized object into a clean high vacuum, such as cavity optomechanics with one or multiple particles to study macroscopic quantum mechanics \cite{RomeroIsart2011Superpositions}, phase transitions \cite{Lechner2013Nanodumbbells}, short range forces \cite{Geraci2010ShortRangeForce}, nanoscale heat transport \cite{chiloyan2015transition}, coherent particle-particle interactions \cite{okamoto2013coherent} and gravitational waves \cite{Arvanitaki2013Gravitational}.\\

\section*{Acknowledgments}
PM, JB and RQ acknowledge financial support from the Fundaci\'o Privada Cellex Barcelona, CoG ERC QnanoMECA (No. 64790), the Spanish Ministry of  Economy and Competitiveness (grant FPU-AP-2012-3729 and FIS2013-46141-P). MS acknowledges financial support from the Ministry of Science and Technology of the Republic of Serbia (project OI171005) and Marie Curie COFUND (FP7-PEOPLE-2010-COFUND).  LN and  JG acknowledge financial  support  from ERC-QMES (no. 338763).



%

\end{document}